\documentclass[referee]{raa}            
\usepackage{graphicx,times}             
\usepackage{epstopdf}
\usepackage{footnote}
\begin{document}

   \title{Is TOL 1326--379 a Prototype of $\gamma$-ray Emitting FR0 Radio Galaxy?}

   \volnopage{Vol.0 (200x) No.0, 000--000}      
   \setcounter{page}{1}          

   \author{Wen-Jing Fu
      \inst{1}
   \and Hai-Ming Zhang
      \inst{2}
   \and Jin Zhang
      \inst{3\dag}
   \and Yun-Feng Liang
      \inst{1}
   \and Su Yao
      \inst{4}
   \and En-Wei Liang
      \inst{1}
   }

   \institute{Guangxi Key Laboratory for Relativistic Astrophysics, School of Physical Science and Technology, Guangxi University, Nanning 530004, People's Republic of China\\
        \and
        School of Astronomy and Space Science, Nanjing University, Nanjing 210023, People's Republic of China
        \and
          School of Physics, Beijing Institute of Technology, Beijing 100081, People's Republic of China; j.zhang@bit.edu.cn
        \and
        Max-Planck-Institute f\"{u}r Radioastronomie, Auf dem H\"{u}gel 69, 53121 Bonn, Germany}

   \date{Received~~201X month day; accepted~~201X~~month day}

\abstract{With the possible spacial association to the Fermi/LAT source 3FGL J1330.0--3818, TOL 1326--379 may be the first one that is identified as a $\gamma$-ray emitting Fanaroff--Riley type 0 radio galaxy (FR0 RG). We analyze the $\sim$12 yr Fermi/LAT observation data of this $\gamma$-ray source and examine its association to TOL 1326--379. We show that the $\gamma$-ray source (named as J1331.0--3818) is tentatively detected with a TS value of 28.7, 3FGL J1330.0--3818 is out of the 95\% containment of J1331.0--3818, and their positions are spatially separated $\sim0.2\degr$. 4FGL J1331.3--3818 falls into the 68\% containment of J1331.0--3818, suggesting that our result agrees with that reported in the Fourth Fermi LAT Source Catalog. TOL 1326--379 is out of the 95\% containment of J1331.0--3818, and their positions are spatially separated $\sim0.4\degr$, indicating that the association between J1331.0--3818 and TOL 1326--379 is quite ambiguous. However, we do not find other possible potential radio or X-ray counterpart within the circle centered at J1331.0--3818 with a radius of 0.4$\degr$. The spectral energy distribution (SED) of TOL 1326--379 shows a bimodal feature as seen in the $\gamma$-ray emitting RGs. We fit the SED with the one-zone leptonic model and find that the average energy spectrum of J1331.0--3818 agrees with the model prediction. Assuming that J1331.0--3818 is an unidentified $\gamma$-ray source, we derive the upper-limit of the $\gamma$-ray flux for TOL 1326--379. It is not tight enough to exclude this possibility with the SED modeling. Based on these results, we cautiously argue that the $\gamma$-ray source J1331.0--3818 is associated with TOL 1326--379 and its jet radiation physic is similar to those $\gamma$-ray emitting RGs.
}

   \authorrunning{Wen-Jing Fu et al. }            
   \titlerunning{Is TOL 1326--379 a Prototype of $\gamma$-ray Emitting FR0 Radio Galaxy? }  

   \maketitle
%
%
\section{Introduction}           
\label{sect:intro}

Radio galaxies (RGs) are a sub-class of the radio-loud active galactic nucleus (AGNs) and are also thought to be the parent populations of blazars with large viewing angles (Urry \& Padovani 1995). Generally, RGs have the large-scale jets extending to kpc--Mpc scale and are respectively classified as Fanaroff--Riley type I (FRI) and II (FRII) RGs according to their radio morphologies, i.e., corresponding to the core-dominated RGs with ``edge-dimmed'' and lobe-dominated RGs with ``edge-brightened'' (Fanaroff \& Riley 1974). On average FRII RGs have the higher radio powers than FRI RGs (Zirbel \& Baum 1995). Based on the optical spectrum properties, RGs could be divided into high excitation galaxies (HEGs) and low excitation galaxies (LEGs;  Hine \& Longair 1979; Laing et al. 1994). FRI RGs and FRII RGs normally correspond to LEGs and HEGs, respectively, however, they do not have the one-to-one correspondence (Hine \& Longair 1979; Laing et al. 1994; Hardcastle et al. 2009b). The two types of RGs may have the intrinsically different accretion modes, which may be also unified with the two kinds of blazars (Ghisellini \& Celotti 2001; Wu \& Cao 2008; Xu et al. 2009; Zhang et al. 2014; Zhang et al. 2015). Due to the remarkable large-scale jets, the communities always focus on the study of these typical RGs (FRI and FRII RGs).

Recently, the cross-match of the Sloan Digital Sky Survey (SDSS), the Faint Images of the Radio Sky at Twenty centimetres survey (FIRST), and the National Radio Astronomy Observatory Very Large Array Sky Survey (NVSS) show that the majority of AGNs at low redshift are compact sources with linear sizes $\leq10$ kpc (Best et al. 2005; Best \& Heckman 2012). These ``compact'' sources were named as ``FR0'' RGs (Ghisellini 2011; Sadler et al. 2014; Baldi et al. 2015) in order to emphasize the distinction with the traditional extended RGs. FR0 RGs have the analogous bolometric luminosities, host galaxy properties, and black hole masses to the low-power LEGs (or FRI RGs, Capetti et al. 2017; Baldi et al. 2019a; Capetti et al. 2020b), except for that the large-scale environment of FR0 RGs may be poorer than that of FRI RGs (Capetti et al. 2020b). Both spectral and morphological features indicate that a fraction of FR0 RGs display the jet structure. Hence FR0 RGs may represent a continuous population of jetted sources at the low end in size and radio power (Capetti et al. 2020a), and the significant Doppler boosting effect has been observed in some FR0 RG pc-scale jets (Cheng \& An 2018). Whether the compact structure without evident extended radio emission of FR0 RGs is due to the different central engine mechanisms from the typical FRI and FRII RGs, or FR0 RGs are on early stage and are the young FRI RGs (e.g., Capetti et al. 2020a), or FR0 RGs are the parent populations of the truly faint low-power BL Lacs (Baldi et al. 2019b), it is unclear.

More than forty FRI/FRII RGs have been detected by the Large Area Telescope on board the Fermi satellite (Fermi/LAT, Abdollahi et al. 2020) and six RGs\footnote{Six TeV RGs are Cen A, M 87, NGC 1275, PKS 0625--35, IC 310, and 3C 264.} are confirmed as TeV $\gamma$-ray emitters. RGs are thought to have the significant contributions to the extragalactic diffuse $\gamma$-ray background (Di Mauro et al. 2014). Moreover, 4\%--18\% of the unresolved $\gamma$-ray background below 50 GeV observed by the Fermi/LAT is likely from the low-luminosity compact core-dominated radio sources, including FR0 RGs and core galaxies (Stecker et al. 2019). However, owing to the low-luminosity, low beaming effect, and steep spectrum at the $\gamma$-ray band, they are difficult to be detected in the GeV-TeV band with the Fermi/LAT and current TeV $\gamma$-ray instruments.

TOL 1326--379 is the first FR0 RG that is tentatively identified as the counterpart of the Fermi/LAT source 3FGL J1330.0--3818 with the possible spacial association in the Third Fermi LAT Source Catalog (3FGL; Acero et al. 2015). 3FGL J1330.0--3818 was detected at a significance of $\sim5\sigma$ with a flux of $F_{\rm1-100 GeV}=3.13\times10^{-10}$ ph cm$^{-2}$ s$^{-1}$ (Acero et al. 2015). The association between 3FGL J1330.0--3818 and TOL 1326--379 was evaluated with the Bayesian method and the derived Bayesian probability of association is 90\% in the third catalog of AGN detected by the Fermi/LAT (3LAC; Ackermann et al. 2015). To further check the 95\% $\gamma$-ray position error ellipse of 3FGL J1330.0--3818, Grandi et al. (2016) suggested that TOL 1326--379 is still the most likely associated low-energy counterpart of 3FGL J1330.0--3818. However, no $\gamma$-ray counterpart for TOL 1326--379 is reported in the Fourth Fermi LAT Source Catalog (4FGL, Abdollahi et al. 2020). 3FGL J1330.0--3818 is the same source of 4FGL J1331.3--3818 in the 4FGL (Abdollahi et al. 2020). No such an association is claimed in the 4FGL since the position of 4FGL J1331.3--3818 shifts away from TOL 1326--379. This makes a confusion whether TOL 1326--379 is the counterpart of the Fermi/LAT $\gamma$-ray source. In addition, the physical classification of TOL 1326--379 is under debated. It locates at the redshift of $z=0.02843$ (Jones et al. 2009) and was proposed as a flat-spectrum radio quasar (FSRQ), an usually seen association between $\gamma$-ray sources and FSRQs in the 3FGL (Acero et al. 2015). Nevertheless, Grandi et al. (2016) suggested that it is an FR0 RG by analyzing its multi-wavelength properties. The detection of $\gamma$-ray emission in FR0 RGs opens up some interesting scenarios on the jet physics (e.g., Baldi et al. 2019b).

In this paper, we intend to study the association between the $\gamma$-ray source and TOL 1326--379 using the $\sim$12 yr Fermi/LAT observation data. We also investigate the possible radiation mechanism of the $\gamma$-rays for TOL 1326--379 and compare its jet properties with other $\gamma$-ray emitting AGNs. The Fermi/LAT data analysis is presented in Section 2, the association analysis is given in Section 3, the discussion about the jet properties is reported in Section 4, and conclusions are given in Section 5. Throughout, $H_{0}$=71 km s$^{-1}$ Mpc$^{-1}$, $\Omega_{m}$=0.27, and  $\Omega_{\Lambda}$= 0.73 are used.

\section{\emph{Fermi}/LAT Data Analysis}

The latest Pass 8 Fermi/LAT data observed from 2008 August to 2020 July (MJD 54682--59035) are chosen in this analysis. The selected region of interest (ROI) is a circle with a radius of 10$^\circ$ centered at the optical position of TOL 1326--379 (R.A.=202.330$\degr$, Decl.=-38.239$\degr$). Events with energies from 0.1 to 300 GeV are selected with a quality criteria of ``$(DATA\_QUAL>0)\&\&(LAT\_CONFIG==1)$''. The standard Fermi Science Tool software package\footnote{https://fermi.gsfc.nasa.gov/ssc/data/analysis/documentation/Cicerone/} (version v11r5p3) and the $P8R3\_SOURCE\_V2$ set of instrument response functions (IRFs) are utilized for the analysis. To minimize the effect of the $\gamma$-ray-bright Earth limb, events with zenith angle larger than 90$^\circ$ are excluded.

Since the 4FGL point sources are based on the 8 yr survey data, we first make the new background source test of the ROI with the \emph{gtfindsrc} package by subtracting the diffuse Galactic emission with the parameterized model of $gll\_iem\_v07.fits$, the isotropic background model of $iso\_P8R3\_V2\_v1.txt$, and the $\gamma$-ray sources listed in the 4FGL (including the source 4FGL J1331.3--3818). The normalization parameters of the diffuse Galactic $\gamma$-ray emission and the isotropic component, and both the normalization and spectral parameters of the point sources are set as free parameters. A binned maximum likelihood analysis is implemented with the \textit{gtlike} tool for optimizing the background model parameters. The maximum likelihood test statistic (TS) is used to estimate the significance of $\gamma$-ray signals, which is defined as TS$=2\log(\frac{\mathcal{L}_{\rm src}}{\mathcal{L}_{\rm null}})$, where $\mathcal{L}_{\rm src}$ and $\mathcal{L}_{\rm null}$ are the likelihoods of the background with or without the point source, respectively. We generate the initial residual TS map for the background of the ROI, and find two new point sources at R.A.=201.120$\degr$, Decl.=-39.347$\degr$ and at R.A.=202.675$\degr$, Decl.=-39.696$\degr$ with TS$>$25. We obtain the current background model by adding the two new sources and generate again the residual TS map for the background. The maximum TS value of the TS residual map is 9.8.

\begin{figure}
 \centering
  \includegraphics[angle=0,scale=0.64]{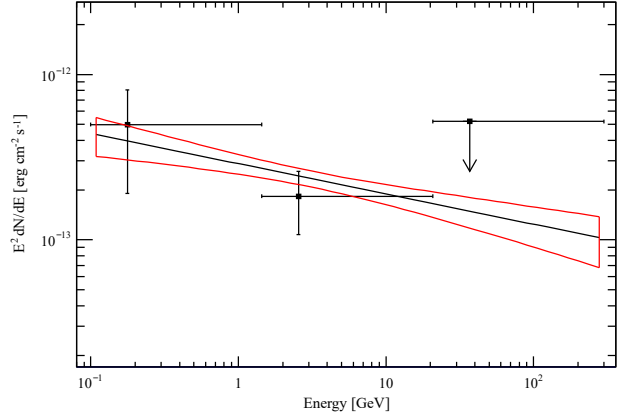}
  \caption{Average energy spectrum of J1331.0--3818 derived with the $\sim$12 yr Fermi/LAT observations in the 0.1--300 GeV band. If TS$<9$, an upper-limit is presented for the energy-bin. The black solid line represents the fitting result with a power-law model and the red bow indicates the 1$\sigma$ uncertainty.}\label{energy-spe}
\end{figure}

Using the current background model (in this case, 4FGL J1331.3--3818 is not subtracted as a background source) and the $\sim$12 yr Fermi/LAT observation data, we utilize the \textit{gtfindsrc} tool to re-estimate the best-fit position of the $\gamma$-ray source with the unbinned likelihood method. To distinguish with 3FGL J1330.0--3818 and 4FGL J1331.3--3818, we name the re-estimated best-fit position of the $\sim$12 yr Fermi/LAT observation data as ``J1331.0--3818'', which is located at R.A.=202.739$\degr$, Decl.=-38.309$\degr$ with an uncertainty radius of 0.08$\degr$ at the 68\% confidence level. The TS value of J1331.0--3818 is 28.7 and the derived TS map with the 68\% and 95\% containment circles is shown in Figure \ref{TSmap}. The derived average energy spectrum of J1331.0--3818 is shown in Figure \ref{energy-spe}. Note that TS=9 approximately corresponds to $\sim3\sigma$ detection (Mattox et al. 1996). Since the total TS value is only 28.7, the spectrum is obtained with three energy bins. All the detected photon energies are lower than 20 GeV. The derived photon spectral index is $\Gamma_{\gamma}=2.18\pm0.20$ and the $\sim$12 yr average flux is $(2.3\pm0.5)\times 10^{-9}$ ph cm$^{-2}$ s$^{-1}$.

\section{Gamma-ray Source Association Analysis}

\begin{figure}
 \centering
  \includegraphics[angle=0,scale=0.25]{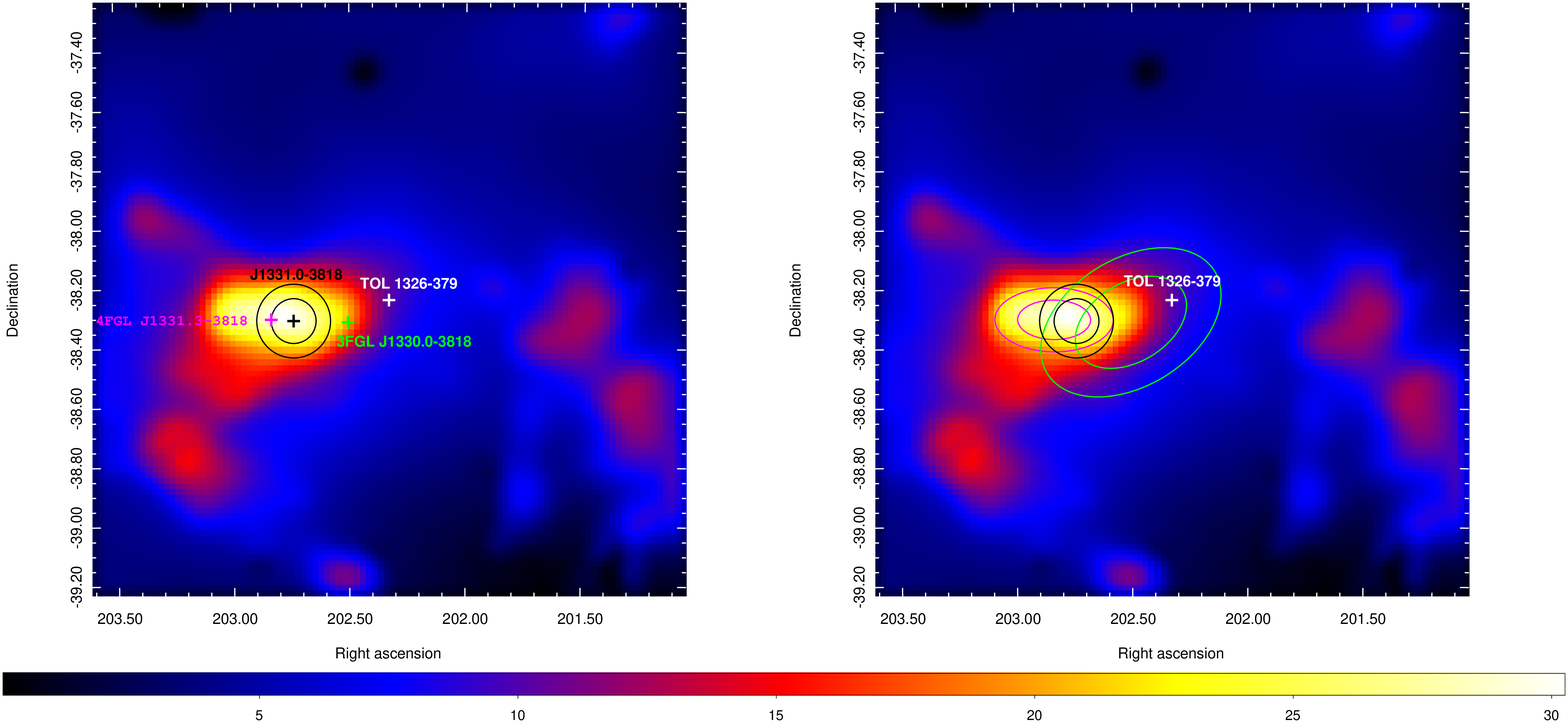}
  \includegraphics[angle=0,scale=0.25]{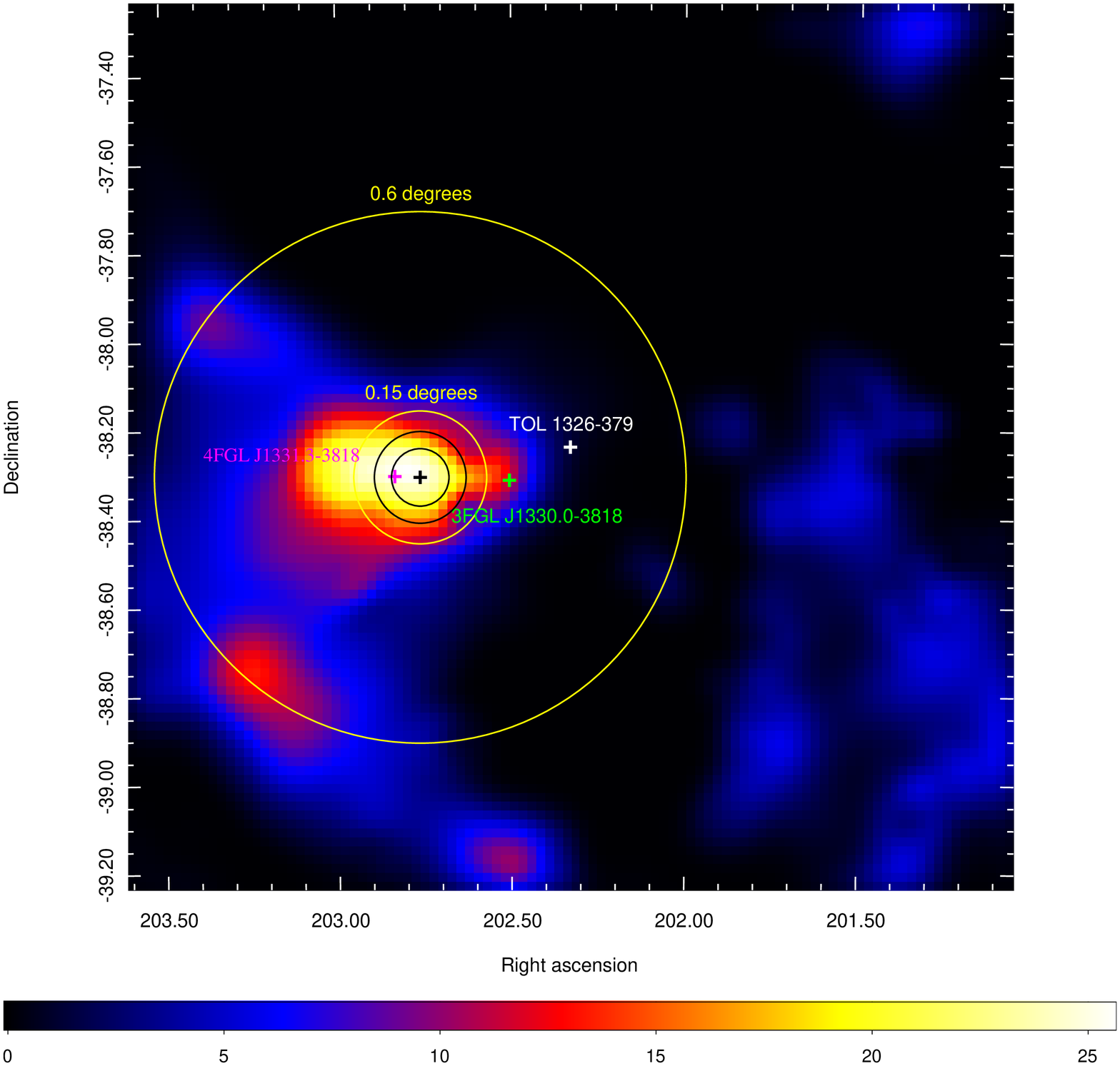}
 \caption{\emph{Top-left Panel}: Smoothed TS map using a Gaussian kernel of 0.12$\degr$ derived with the $\sim$12 yr Fermi/LAT observation data in the 0.1--300 GeV band. The TS map, centered at the optical position of TOL 1326--379 (white cross), is $2^\circ\times2^\circ$ with a pixel size of 0.02 degrees/pixel. The best-fit position (named as J1331.0--3818) is marked as a black cross and the corresponding 68\% and 95\% containments are marked with black circles. The green and magenta crosses respectively represent the positions of 3FGL J1330.0--3818 and 4FGL J1331.3--3818. \emph{Top-right Panel:} Same as the \emph{top-left panel}, but the 68\% and 95\% containments of the positions for 4FGL J1331.3--3818 (magenta ellipses) and 3FGL J1330.0--3818 (green ellipses) are presented for comparison. \emph{Bottom Panel:} Same as the \emph{top-left panel}, but in the 1--300 GeV band. The yellow circles represent the 68\% containment of the Fermi/LAT PSF at 1 GeV ($0.6^{\circ}$) and 10 GeV ($0.15^{\circ}$, Atwood et al. 2009), which is centered at the position of J1331.0--3818.
}\label{TSmap}
\end{figure}

\begin{figure}
  \centering
  \includegraphics[angle=0,scale=1.]{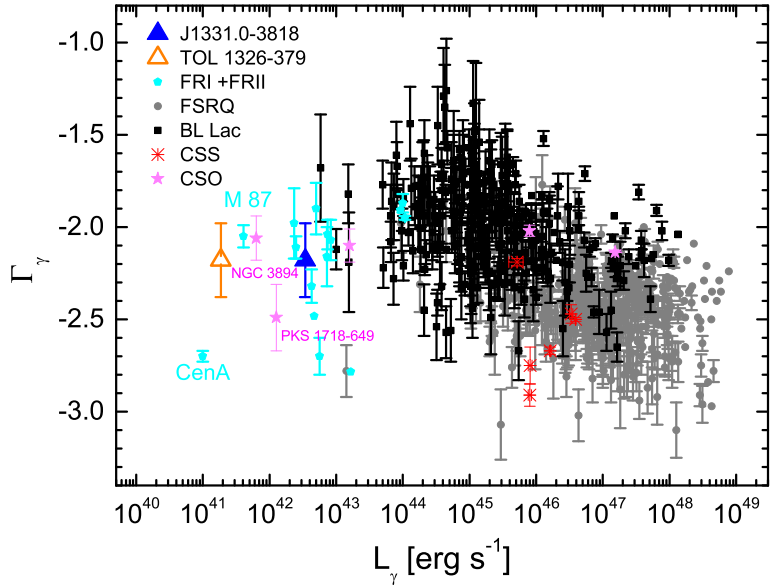}
  \caption{$\Gamma_{\gamma}$ vs. $L_{\gamma}$. The average luminosity of J1331.0--3818 and the luminosity upper-limit of TOL 1326--379 are shown as blue solid and orange opened triangles, respectively. The data of blazars and RGs are taken from the 3FGL. The CSO data are from Abdollahi et al. (2020) and Gan et al. (2021) while the compact steep-spectrum source (CSS) data are taken from Zhang et al. (2020).}\label{gamma-L}
\end{figure}

To compare our results with that in the 3FGL and 4FGL, the positions of 3FGL J1330.0--3818 and 4FGL J1331.3--3818 are also shown in Figure \ref{TSmap}. It is found that they are not completely overlapped. 4FGL J1331.3--3818 derived from the 8 yr Pass 8 Fermi/LAT data marginally falls into the 68\% containment circle of the source J1331.0--3818 while 3FGL J1330.0--3818 derived from the 4 yr Fermi/LAT data is out of the 95\% containment circle of J1331.0--3818. As shown in Figure \ref{TSmap}, the 68\% and 95\% containments of 3FGL J1330.0--3818 are accordingly much broader than that of 4FGL J1331.3--3818 and J1331.0--3818. Note that TS=29.6 was yielded for 3FGL J1330.0--3818 with the 4 yr Fermi/LAT data in the 3FGL (Acero et al. 2015) and TS=17.7 was given for 4FGL J1331.3--3818 with 8 yr Pass 8 Fermi/LAT data in the 4FGL (Abdollahi et al. 2020). Using the $\sim$12 yr Pass 8 Fermi/LAT data, we obtain TS=28.7 for J1331.0--3818 in this analysis, and the source position constraint is roughly consistent with that of 4FGL J1331.3--3818, and is tighter than 3FGL J1330.0--3818. We compare J1331.0--3818 with other Fermi/LAT detected AGNs in the photon spectral index ($\Gamma_{\gamma}$) versus $\gamma$-ray luminosity ($L_{\gamma}$) plane in Figure \ref{gamma-L}. It locates at the typical RG area, not the blazar region, and almost occupies the low luminosity end. The luminosities of two typical RGs (M 87 and Cen A) and two compact symmetric objects (CSOs, NGC 3894 and PKS 1718--649) are much lower than the $\sim$12 yr average luminosity of J1331.0--3818.

The optical position of TOL 1326--379 is also shown in Figure \ref{TSmap}. It is in the 68\% error ellipse of 3FGL J1330.0--3818 (Figure \ref{TSmap}), and the association between TOL 1326--379 and 3FGL J1330.0--3818 was claimed in the 3FGL (Acero et al. 2015). Grandi et al. (2016) reported that there are other three bright radio sources at 1.4 GHz within the 95\% $\gamma$-ray position error ellipse of 3FGL J1330.0--3818, but they are not detected in the 4.8 GHz radio survey with the PARKES telescope (Griffith \& Wright 1993) and the Two Micron All Sky Survey (2MASS, Skrutskie et al. 2006). The radio source associated with TOL 1326--379 has a flat spectrum with an index of $\alpha=0.37$ in the 843 MHz -- 20 GHz band and has a 2MASS counterpart. Therefore, they suggested that TOL 1326--379 is the most likely association of 3FGL J1330.0--3818. However, TOL 1326--379 is convincingly out of the 95\% containment of 4FGL J1331.3--3818, and no association between them was reported in the 4FGL (Abdollahi et al. 2020).

Our result agrees with that reported in the 4FGL. TOL 1326--379 is 0.4$^\circ$ away from the position of J1331.0--3818. Note that the angular resolution of the Fermi/LAT increases with the energy band from $<3.5^{\circ}$ at 100 MeV to $<0.15^{\circ}$ at $>$10 GeV (Atwood et al. 2009). Thus, we further examine the position of J1331.0--3818 with its high energy photons. Our first try is to use the $\gamma$-ray photons above 10 GeV, but we cannot find out the source, indicating that the $\gamma$-ray photons of J1331.0--3818 are mostly below 10 GeV. It is consistent with that shown in Figure \ref{energy-spe}. Thus, we use the $\gamma$-ray photons above 1 GeV for this purpose. The TS map in the energy band of 1--300 GeV is also shown in Figure \ref{TSmap}. The TS value is 25.8 and the flux in the energy band is (1.6 $\pm$ 0.5) $\times$ 10$^{-10}$ ph cm$^{-2}$ s$^{-1}$. The best-fit position is consistent with that derived by the $\gamma$-ray photons in the 0.1--300 GeV band within the error bar. The 68\% containment circles of the point spread function (PSF) for the Fermi/LAT at 1 GeV ($R=0.15^{\circ}$) and at 10 GeV ($R=0.6^{\circ}$, Atwood et al. 2009) are also presented in Figure \ref{TSmap}. The positions of 3FGL J1330.0--3818, 4FGL 1331.3--3818, and J1331.0--3818 all fall into the error circle of $R=0.6^{\circ}$. However, the lack of the photons with energy $>$10 GeV makes it be difficult to absolutely resolve their locations. TOL 1326--379 is out of the error circle of $R=0.15^{\circ}$, but also in the error circle of $R=0.6^{\circ}$. One still cannot exclude the possible spatial association between J1331.0--3818 and TOL 1326--379 with the Fermi/LAT observations.

We hence search the likely potential radio or X-ray counterparts of the $\gamma$-ray source J1331.0--3818 within a circle, which is centered at the position of J1331.0--3818 with an angular radius of 0.4$^\circ$. We find 13 objects within this area circle in the Sydney University Molonglo Sky Survey (SUMSS; Mauch et al. 2003) at 843 MHz, including TOL 1326--379, as listed in Table 1. We then check these radio sources at higher frequency survey and do not find any radio sources, including TOL 1326--379, in the 0.08--22 GHz Parkes Catalog (Wright \& Otrupcek 1990). In the 4.85 GHz Parkes MIT-NRAO Catalog (PMN; Gregory et al. 1994), besides TOL 1326--379, two other radio sources are returned, as listed in Table 2. However, no observation data at other bands are presented in the NASA/IPAC Extragalactic Database (NED) for the two radio sources. And only TOL 1326--379 is listed in the Combined Radio All-Sky Targeted Eight GHz Survey (CRATES; Healey et al. 2007). In the X-ray band, the Swift detects an X-ray source in the 15-150 keV band at R.A.=202.330$\degr$, Decl.=-38.239$\degr$ (Evans et al. 2020), and ROSAT also detects this source in the 0.1-3 keV band at R.A.=202.334$\degr$, Decl.=-38.243$\degr$ (Boller et al. 2016). This X-ray source is associated with TOL 1326--379. Hence, TOL 1326--379 is the most possible $\gamma$-ray emitting counterpart within this area circle.

Although TOL 1326--379 is the most possible counterpart of the $\gamma$-ray source, one still cannot exclude the possibility that the $\gamma$-ray source is an unidentified source due to the large offset of the positions between TOL 1326--379 and the $\gamma$-ray source. In this scenario, we place an upper-limit of the $\gamma$-ray emission for TOL 1326--379 by adding the $\gamma$-ray source 4FGL J1331.3--3818 as a background source. We obtain $F_{\rm limit}=8.17 \times10^{-10}$ ph cm$^{-2}$ s$^{-1}$.\

\section{Discussion}

\subsection{Optical Variability and Implication for the Type of TOL 1326--379}

TOL 1326--379 was reported as a point source in the CRATES and classified as an FSRQ (Healey et al. 2007). Whereas Grandi et al. (2016) suggested that it is an FR0 RG by analyzing its multi-wavelength properties. In the radio images, no extended emission is detected for TOL 1326--379. In the $L_{\rm 1.4 GHz}-L_{[\rm O~{\scriptsize III}]}$ plane, it falls into the region of typical FR0 RGs. Its optical spectrum is also shown as a canonical LEG spectrum and demonstrates the against characteristics with FSRQs.

\begin{figure}
  \centering
  \includegraphics[angle=0,scale=0.8]{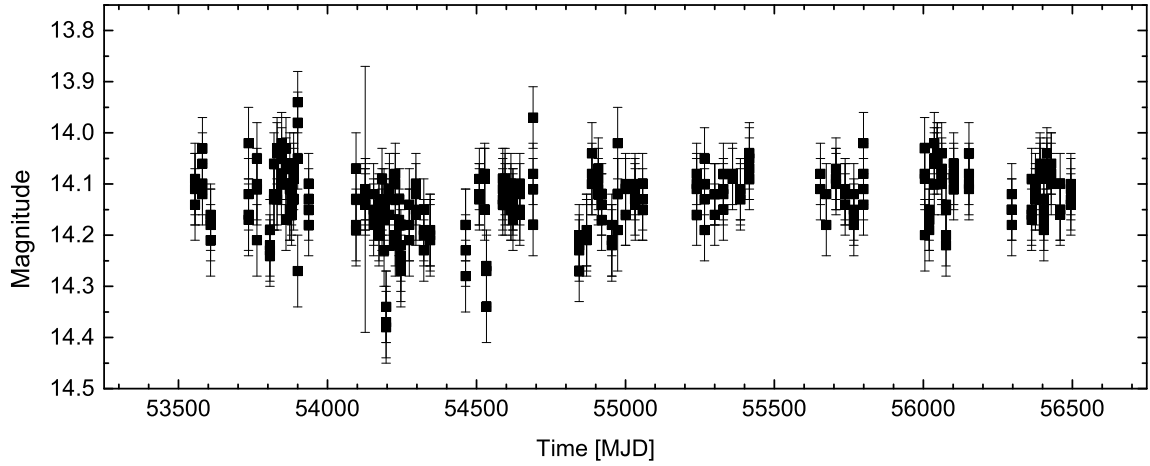}
  \caption{Optical V-band light curve of TOL 1326--379. The data are taken from the CRTS.}\label{LC}
\end{figure}

It is well known that FSRQs usually show significant variability in multi-wavelengths. We examine the optical variability of TOL 1326--379 with a long-term monitoring project by the Catalina Real-Time Transient Survey (CRTS; http://crts.caltech.edu/;  Drake et al. 2009; Mahabal et al. 2011). Figure \ref{LC} displays its V-band light curve, which covers about 3000 days from MJD 53607 (Aug. 25, 2005) to MJD 55416 (Aug. 08, 2010). We measure the flux variability with $\Delta V=\frac{1}{N}\sum_i(|V_i-\bar{V}|/\sigma_{V_i})$, where N is the number of the data points, $\bar{V}$ is the average magnitude, and $\sigma_{V_i}$ is the error of the magnitude $V_i$ for data point $i$. In case of $\Delta V>3$, the global variation is statistically claimed at a significance lever of 3$\sigma$. We obtain $\Delta V=0.76$, indicating that TOL 1326--379 is almost quiet in this time period. This disfavors TOL 1326--379 as an FSRQ in the variability point of view.

\subsection{X-ray Radiation Physics of TOL 1326--379 and Implication for its $\gamma$-ray Origin}

TOL 1326--379 is bright in the X-ray band (Torresi et al. 2018). We compile its SED in the radio-optical-X-ray band to study the radiation mechanism of X-rays and explore the origin of $\gamma$-rays. The SED data are from the literature (Tavecchio et al. 2018), as illustrated in Figure \ref{sed}. One can observe that the X-rays do not follow the extrapolation of the radio-to-optical component and shape as a hard spectrum component. Such a bimodal feature is usually seen in the SEDs of blazars and $\gamma$-ray emitting RGs (Abdo et al. 2009; Migliori et al. 2011; Zhang et al. 2012; Zhang et al. 2014; Fukazawa et al. 2015; Xue et al. 2017). Thus, we represent the SED with the one-zone leptonic model, in which the bump in the radio-optical band is attributed to the synchrotron radiation of relativistic electrons and the high energy bump is explained with the synchrotron self-Compton (SSC) scattering process (see also Tavecchio et al. 2018). Note that the three UV data points are contaminated by the host galaxy emission (Grandi et al. 2016). We take them as upper-limits only to constrain the model parameters during the SED fits. In the SED modeling, the emitting region is assumed as a sphere with radius $R$ and magnetic field $B$. It has relativistic motion with a Doppler factor $\delta$, where $\delta=1/(\Gamma-\sqrt{\Gamma^2-1}\cos\theta)$, $\Gamma$ and $\theta$ are the bulk Lorenz factor and viewing angle of the radiation region. The electron distribution is taken as a broken power-law,
\begin{equation}
N(\gamma )= N_{0}\left\{ \begin{array}{ll}
\gamma ^{-p_1}  &  \mbox{ $\gamma_{\rm min}\leq\gamma \leq \gamma _{\rm b}$}, \\
\gamma _{\rm b}^{p_2-p_1} \gamma ^{-p_2}  &  \mbox{ $\gamma _{\rm b} <\gamma\leq\gamma _{\rm max} $.}
\end{array}
\right.
\end{equation}

During the SED modelling, the Klein--Nishina effect and the absorption of high-energy $\gamma$-ray photons by the extragalactic background light (Franceschini et al. 2008) are also taken into account. Note that the low-frequency radiation below 10 GHz is usually thought to be from the extended region due to the synchrotron-self-absorption effect in the core region. Therefore, we do not take the data at $\nu<10$ GHz into account. The indices $p_{1}$ and $p_{2}$ can be derived from the spectral indices by fitting the observed data with a power-law function (see also Zhang et al. 2012; Xue et al. 2017). $\gamma_{\min}$ and $\gamma_{\max}$ are generally poorly constrained. The $R$ value is normally estimated with the variability timescale ($\Delta t$), i.e., $R=\delta\Delta t c/(1+z)$. $B$ and $\delta$ are degenerate (e.g., Zhang et al. 2012). With the current SED data, the model parameters cannot be constrained. We fix $R=5\times10^{16}$ cm, $p_1=2.0$, $\gamma_{\min}=100$, $\gamma_{\max}=3\times10^{4}$, $\theta=30\degr$, as did in Tavecchio et al. (2018). And then we find that the SED in the optical-X-ray band can be well represented with the model by taking the rest parameters as follows, $B=1.5$ G, $\Gamma=1.05$ ($\delta=1.3$), $p_2=4.1$, $\gamma_{\rm b}=2500$, and $N_0=3300$ cm$^{-3}$. The model curve is shown in Figure \ref{sed}. For comparison, the result reported by Tavecchio et al. (2018) is also presented.

We add the average energy spectrum of the $\gamma$-ray source J1331.0--3818 to Figure \ref{sed}. Interestingly, its $\gamma$-rays agree with the model prediction, indicating that the $\gamma$-rays are likely associated with TOL 1326--379. Assuming that J1331.0--3818 is an unidentified $\gamma$-ray source, we derived the upper-limit of the $\gamma$-rays for TOL 1326--379 in Section 3 and add the upper-limit in Figure \ref{sed}. We cannot also exclude this possibility as displayed in Figure \ref{sed}. For comparison, the fitting result and parameters with the same model in Tavecchio et al. (2018) are also given in Figure \ref{sed} and Table 3. Note that the $\gamma$-ray fluxes used in Tavecchio et al. (2018), which are taken from the 3FGL, are much higher than that of J1331.0--3818 derived in Section 2. The bimodal SED in our result is more dominated by the synchrotron radiation, and thus a larger $B$ and a smaller $\gamma_{\rm b}$ is obtained than that in Tavecchio et al. (2018) for the same synchrotron radiation peak frequency.

\section{Conclusions}

We analyzed the $\sim$12 yr Fermi/LAT observation data of 3FGL J1330.0--3818/4FGL J1331.3--3818 for exploring its association with TOL 1326--379, and the re-estimated best-fit position with the \emph{gtfindsrc} tool is named as J1331.0--3818 for distinguishing with 3FGL J1330.0--3818/4FGL J1331.3--3818. The $\gamma$-ray source J1331.0--3818 is detected with a TS value of 28.7 and the $\sim$12 yr average flux is $(2.3\pm0.5)\times10^{-9}$ ph cm$^{-2}$ s$^{-1}$ with a photon spectral index of $\Gamma_{\gamma}=2.18\pm0.20$. The positions of J1331.0--3818, 3FGL J1330.0--3818, and 4FGL J1331.3--3818 are not overlapped. The position of 4FGL J1331.3--3818 falls into the 68\% containment of J1331.0--3818 in the derived TS map while 3FGL J1330.0--3818 is out of the 95\% containment and is spatially separated $0.2\degr$ from J1331.0--3818. TOL 1326--379 is out of the 95\% containment of J1331.0--3818, and their positions are spatially separated $0.4\degr$. Since the spatial association between TOL 1326--379 and J1331.0--3818 is quite uncertain, we try to find out any likely associated candidates of J1331.0--3818 by cross-checking the radio and X-ray survey catalogs. We did not find out other possible counterparts. We also checked the optical flux variation of TOL 1326--379. This source is quiet and no significant outburst is found during 2005-2010, indicating that it is not similar to blazars. The compiled SED of TOL 1326--379 in the radio-optical-X-ray band shows a bimodal feature as seen in $\gamma$-ray emitting RGs. We fit the SED with the one-zone leptonic model and find that the derived average energy spectrum of J1331.0--3818 agrees with the model prediction. Assuming that J1331.0--3818 is an unidentified $\gamma$-ray source, we derived the upper-limit of the $\gamma$-ray flux for TOL 1326--379. We cannot also exclude this possibility by the SED fitting. Based on these analysis results, we cautiously argue that J1331.0--3818 is likely associated with TOL 1326--379 and its jet radiation physic is similar to those typical $\gamma$-ray emitting RGs.

The SED modeling of TOL 1326--379 indicates that its jet radiation physic is similar to the $\gamma$-ray emitting RGs (e.g., Xue et al. 2017) and BL Lacs (e.g., Zhang et al. 2012). The fitting parameter values of $B$, $\gamma_{\rm b}$, and $\delta$ are also similar to those of typical RGs with on average larger $B$ and smaller $\gamma_{\rm b}$ and $\delta$. $\delta\sim1.3$ of TOL 1326--379 indicates the very weak Doppler boosting effect of its core region. The radiations from X-ray to $\gamma$-ray bands for TOL 1326--379 can be explained with the SSC process. As reported by Baldi et al. (2019b), the Compton component peaks of FR0 RGs likely fall in the MeV band. Therefore, the MeV measurements together with plenty of multiwavelength data will be help to ascertain the peaks in the broadband SEDs and then to explore the GeV--TeV radiation properties of FR0 RGs.

On the basis of the fitting parameters, we also estimate the jet power ($P_{\rm jet}$) of TOL 1326--379. It is assumed that the jet power is carried by the powers of radiation electrons, magnetic fields, and radiations, i.e., $P_{\rm jet}=\sum_{i}P_{i}$, where $P_{i}$ are the powers of relativistic electrons ($P_{\rm e}$), magnetic fields ($P_{B}$), and radiations ($P_{\rm r}$), respectively. Hence we obtain $P_{\rm e}=2.41\times10^{42}$ erg s$^{-1}$, $P_{B}=2.32\times10^{43}$ erg s$^{-1}$, $P_{\rm r}=9.31\times10^{42}$ erg s$^{-1}$, $P_{\rm jet}=3.50\times10^{43}$ erg s$^{-1}$, $P_{\rm r}/P_{\rm jet}=0.27$, and $P_{B}/P_{\rm jet}=0.66$, indicating a magnetized jet with high radiation efficiency in TOL 1326--379, similar to some typical RGs (see Xue et al. 2017).

\begin{figure}
  \centering
  \includegraphics[angle=0,scale=1.]{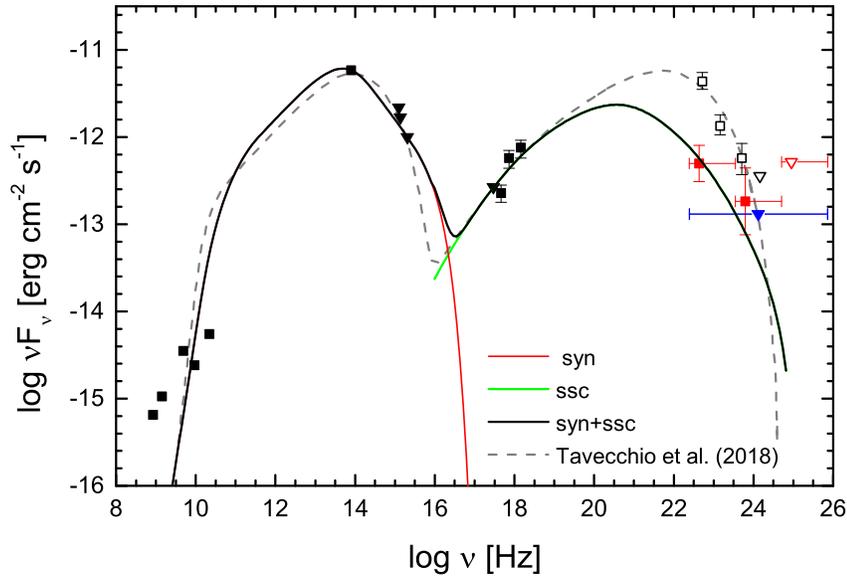}
  \caption{Observed SED with the fitting result by the one-zone leptonic model (black solid line) for TOL 1326--379. The data points marked as black symbols together with the fitting line (gray dashed line) are from Tavecchio et al. (2018), where the opened black symbols are the Fermi/LAT data from the 3FGL and the inverted triangles represent the upper-limits. The red symbols are the $\sim$12 yr average energy spectrum of J1331.0--3818 (same as in Figure 2) while the blue inverted triangle indicates the derived flux upper-limit of TOL 1326--379. }\label{sed}
\end{figure}

\begin{figure}
 \centering
   \includegraphics[angle=0,scale=1.0]{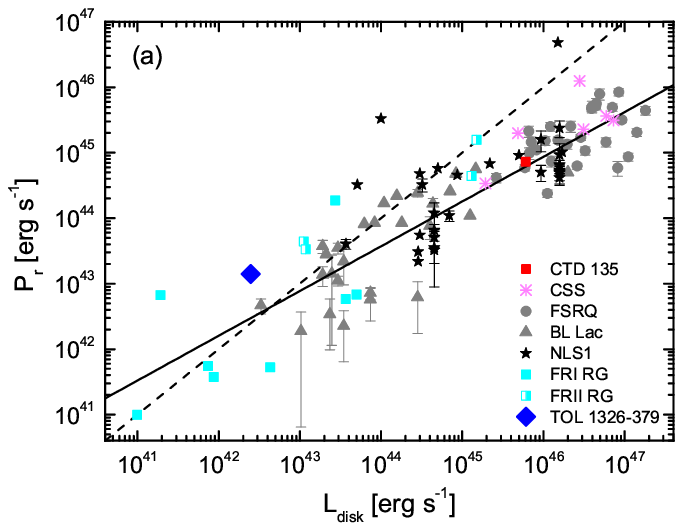}
   \includegraphics[angle=0,scale=1.0]{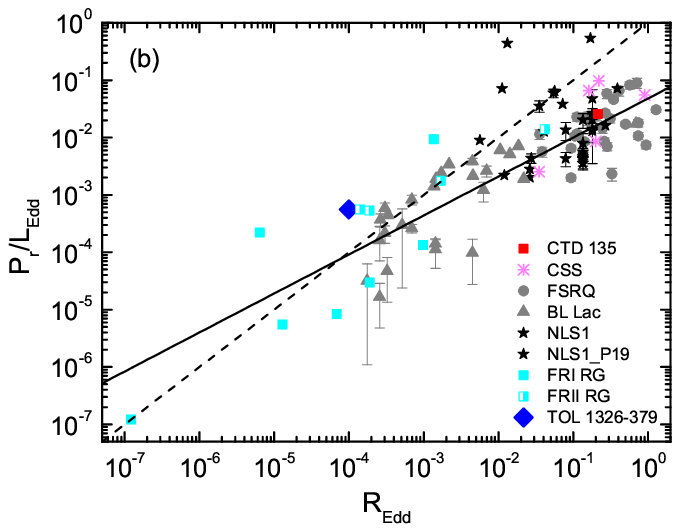}
 \caption{$P_{\rm r}$ as a function of $L_{\rm disk}$ together with their relations in units of Eddington luminosity. The dashed lines indicate the equality line while the solid lines are the linear regression fits for other $\gamma$-ray emitting AGNs (including blazars, narrow-line Seyfert 1
galaxies (NLS1s), RGs, and CSSs, taken from Zhang et al. 2020 and references therein). The data of CTD 135 are from Gan et al. (2021).}\label{Pr-Ldisk}
\end{figure}

As illustrated in Figure \ref{sed}, no obvious thermal-radiation component from the accretion disk is presented in the broadband SED of TOL 1326--379, implying that the disk emission may be overwhelmed by the non-thermal jet radiation. We therefore use the luminosity at $10^{15}$ Hz of the model-fitting line as an upper-limit of the disk luminosity ($L_{\rm disk}$). The derived upper-limit of Eddington ratio for TOL 1326--379 is $R_{\rm Edd}=L_{\rm disk}/L_{\rm Edd}\sim10^{-4}$, by adopting its central black hole mass as $M_{\rm BH}=2\times10^8M_{\bigodot}$ (Grandi et al. 2016), where $L_{\rm Edd}$ is the Eddington luminosity. The Eddington ratio of TOL 1326--379 is smaller than the values of most LEGs (Best \& Heckman 2012), implying a radiation inefficient disk in TOL 1326--379. Using the luminosity of [O~{\scriptsize III}] emission line to estimate the bolometric luminosity, Grandi et al. (2016) obtained $R_{\rm Edd}\sim5\times10^{-3}$, which is roughly coincident with ours. To further compare its jet radiation properties with other $\gamma$-ray emitting AGNs, we put TOL 1326--379 into the $L_{\rm disk}-P_{\rm r}$ and $R_{\rm Edd}-P_{\rm r}/L_{\rm Edd}$ planes of other $\gamma$-ray emitting AGNs (the data to see Zhang et al. 2020, Gan et al. 2021, and references therein). As displayed in Figure \ref{Pr-Ldisk}, TOL 1326--379 falls into the typical RG area and also follows the linear regression fit line of other subclasses of $\gamma$-ray emitting AGNs in Zhang et al. (2020). Hence, its jet radiation may be also connected with the Eddington ratio analogous to other subclasses of AGNs.

\begin{acknowledgements}
This work is supported by the National Natural Science Foundation of China (grants 12022305, 11973050, U1738136, U1731239, 11851304, and 11533003), and Guangxi Science Foundation (grants 2017AD22006, 2019AC20334, and 2018GXNSFGA281007).

\end{acknowledgements}


\clearpage

\begin{table}
\renewcommand{\thetable}{\arabic{table}}
\centering
\caption{Objects within the Circle Centered on J1331.0--3818 with a Radius of 0.4$\degr$ in the SUMSS }
\centering
\label{objects}
\footnotesize
\begin{tabular}{lccccc}
\hline\hline
SUMSS Name & Ra           & Dec          & Int\_flux\_36\_cm & Int\_flux\_36\_cm\_error & Search\_Offset \\
     & [ICRS(2000)] & [ICRS(2000)] & [mJy]             & [mJy]                    & [arc minute] \\
\hline
J133057-381450 & 13 30 57.22 & -38 14 50.1 & 15.5 & 2.1 & 3.675\\		
J133116-381454 & 13 31 16.31 & -38 14 54.7 & 14.7 & 2.1 & 5.175\\	
J133100-380926 & 13 31 00.29 & -38 09 26.1 & 34.8 & 2.3 & 9.093\\
J133158-381411 & 13 31 58.37 & -38 14 11.2 & 69.4 & 2.6 & 12.731\\		
J132952-381249 & 13 29 52.93 & -38 12 49.0 & 102.4 & 7.2 & 13.870\\	
J133227-380959 & 13 32 27.66 & -38 09 59.0 & 21.4 & 2.8 & 19.675\\		
J132919-381416 (TOL 1326-379) & 13 29 19.23 & -38 14 16.2 & 76.5 & 2.9 & 19.721\\		
J133011-383625 & 13 30 11.59 & -38 36 25.9 & 16.9 & 3.4 & 20.037\\		
J133216-383347 & 13 32 16.20 & -38 33 47.1 & 31.3 & 3 & 21.718\\	
J132904-381656 & 13 29 04.66 & -38 16 56.0 & 15.5 & 1.8 & 22.169\\		
J132950-380015 & 13 29 50.52 & -38 00 15.1 & 39.3 & 2.5 & 22.495\\		
J132910-380917 & 13 29 10.99 & -38 09 17.5 & 366.8 & 11.2 & 22.832\\	
J132857-381702 & 13 28 57.33 & -38 17 02.9 & 113.1 & 3.8 & 23.596\\
\hline
\end{tabular}
\end{table}

\begin{table}
\renewcommand{\thetable}{\arabic{table}}
\centering
\caption{Objects within the Circle Centered on J1331.0--3818 with a Radius of 0.4$\degr$ in the PMN Catalog}
\centering
\footnotesize
\begin{tabular}{lccccc}
\hline\hline
PMN Name & Ra        & Dec        & Flux\_4850\_MHz & Flux\_4850\_MHz\_error & Search\_Offset\\
         & ICRS(2000)& ICRS(2000) & [mJy]           & [mJy]                  & [arc minute] \\
\hline
J1330-3812 & 13 30 01.70 & -38 12 47.00 & 55.0 & 9.0 & 12.336\\		
J1329-3815(TOL 1326-379) & 13 29 18.40 & -38 15 02.00 & 66.0 & 10.0 & 19.730\\	
J1329-3811 & 13 29 13.00 & -38 11 10.00 & 98.0 & 10.0 & 21.776\\
\hline
\end{tabular}
\end{table}

\begin{table}
\renewcommand{\thetable}{\arabic{table}}
\centering
\caption{Parameters in SED Modeling of TOL 1326--379}
\centering
\footnotesize
\resizebox{\textwidth}{10mm}{
\begin{tabular}{lccccccccccccccc}
\hline\hline
$R$ & $B$ & $\gamma_{\rm min}$ & $\gamma_{\rm b}$ & $\gamma_{\rm max}$ & $p_{1}$ & $p_{2}$ & $\Gamma$ & $\delta$ & $\theta$ & N$\rm{_{0}}$ & P$\rm{_{e}}$ & P$_{B}$ & P$\rm{_{r}}$ & P$\rm{_{p}}$ & P$\rm{_{jet}}$ \\
{[cm]} & [G] & & & & & & & & [degree] & [cm$^{-3}$] & [erg s$^{-1}$] & [erg s$^{-1}$] & [erg s$^{-1}$] & [erg s$^{-1}$] & [erg s$^{-1}$]\\
\hline
5$\times10^{16}$ & 1.5 & 100 & 2500 & 3$\times10^{4}$ & 2.0 & 4.1 & 1.05 & 1.3 & 30 & 3300 & 2.41$\times10^{42}$ & 2.32$\times10^{43}$ & 9.31$\times10^{42}$ & -- & 3.50$\times10^{43}$ \\
$^a$ 5$\times10^{16}$ & 0.2 & 100 & 10$^{4}$ &  3$\times10^{4}$ & 2 & 4.8 & 2 & 2 & 30 & -- & 3.5$\times10^{43}$ & 1.5$\times10^{42}$ & -- & 1.2$\times10^{44}$ & 1.6$\times10^{44}$\\
\hline
\footnotesize{$^a$ Taken from Tavecchio et al. (2018).}
\end{tabular}}
\end{table}

\end{document}